# Fluctuation in Peer-to-Peer Networks: Mitigating Its Effect on DHT Performance

Dietrich Fahrenholtz, Volker Turau
Telematics Group, Hamburg University of Technology
Schwarzenbergstraße 95
21073 Hamburg
{fahrenholtz,turau}@tuhh.de

**Abstract:** Due to the transient nature of peers, any Peer-to-Peer network is in peril to falling apart if peers do not receive routing table updates periodically. To this end, maintenance, which affects every peer, ensures connectedness and sustained data operation performance. However, a high rate of change in peer population usually incurs lots of network maintenance messages and can severely degrade overall performance. We discuss three methods how to tackle and mitigate the effect of peer fluctuation on a tree-based distributed hash table.

## 1 Introduction

In a vast network of peers without any structure it is hard to find data efficiently because direct routes are frequently not known. Overlay networks, basis of distributed hash tables (DHT) and built when peers join or operate in the network, assist with this kind of search. Numerous DHT protocols have been developed and analysed in terms of number of messages sent or intermediate peers visited per lookup operation. On the underlying DHT, every peer has its own view, which consists of paths to a few other peers. Due to the transient nature of peers, i.e., their lifetime in a Peer-to-Peer network (P2PN) is usually short, this view is in flux and needs to be refreshed periodically. So every peer has to perform this *maintenance* that requires a fraction of the available bandwidth. The more peers come and go at a time the greater this fraction. If the rate of change becomes too high, overall performance may suffer dramatically. This paper shows enhancements to our tree-based DHT to better cope with fluctuation in peer population.

## 2 Characteristics and impact of fluctuation

We understand by *fluctuation* the rate at which peers join or leave the P2PN, while other authors use the term "churn" to refer to the same phenomenon [RGRK03, LSG+04]. However, we find "fluctuation" more suitable as it is about motion like that of waves not like stirring milk to produce butter. The period between a node's joining and its leave from the P2PN is generally called *session time*. In file sharing P2PN such as Gnutella or KaZaa, median session times of merely a few minutes have been witnessed. We believe session times are highly application domain dependent because in one domain requests may simply check for or update a particular datum, whereas in others more lengthy, staged requests need to be performed. Connections made by peers to the P2PN follow a 24-hour pattern [PF95], i.e., there are times of day when many peers arrive and times when only a few join. Furthermore, as was pointed out by [RGRK03], P2PNs that need to store data reliably are more likely to suffer from high fluctuation than those who only need to store pointers to data. This is because required maintenance traffic might quickly exceed the individual bandwidth of a considerable fraction of peers. Since fluctuation in a population of unstable peers is unavoidable, any P2PN needs to grapple with it. That means fluctuation is the natural antagonist of good sustained performance and its effect on the latter needs to be estimated and kept to a minimum.

## 3 A tree-based DHT

In this section, we will roughly sketch our approach [FT04] to a resilient DHT. Due to space constraints, we will only give a short overview.
Every datum to be inserted into the P2PN is stored in peers' memory, which we believe is a cheaper resource than bandwidth. Saroiu et al. state that still at least 35% of all Napster peers report bandwidth of 128Kbps or less [SGG02]. Peers join or leave the P2PN constantly, so the network is always in flux. This inevitably incurs network maintenance, which helps the network stay connected and consistent. So every peer needs to receive new routing table entries from time to time. To find a datum in the P2PN, a unique association with a search key, $k$, (comprising of up to $l$ bits and $k \in [0 \ldots 2^{l-1}]$) is generated. It is derived by applying a secure message digest to the datum. Thus the probability of a collision of two search keys is negligible. During the start phase, there is but one *search key range* (also termed *interval*) from $[0 \ldots 2^{l-1}]$. When a peer, initially empty, joins the P2PN, it assumes responsibility for data of one and only one search key range, i.e., during the join phase it receives a copy of the complete set of data already stored in neighbors' memory. Peers sharing data of a common key range are *neighbors* of one another. Since peers can leave the P2PN or suffer a crash-failure, a minimum number of active peers for each key range warrant for data availability. Active peers in turn are responsive to *insert*, *lookup*, and *update* operations. If an active peer is queried for a datum it actually stores, the peer returns it. An interval needs a *split* into two halves if threshold of number of peers, $g_s$, is exceeded during a growth phase. The left sub-range whose search keys are less than $2^{j-1}$ is called "l", the right sub-range whose search keys are greater than or equal $2^{j-1}$ is called "r", where $2^j$ is the upper bound of the original range and $j < l$. Likewise, if the number of peers in any interval falls below threshold, $g_c$, a *coalesce* operation must be performed. Here "r" and "l" merge to form a super interval. That also means, data formerly belonging to interval "r", for instance, must be replicated to those peers of interval "l" and vice versa. Routing tables (RT) store pointers to other peers. To a selected few of these, each peer is maintaining virtual connections. The level, $d$, in a RT is determined by the size of the interval. Letters $r$ and $l$ in the circles of Fig. 1 denote the RT side of each half, whereas the number next to a letter specifies the level the interval is on. A dashed line between two circles indicates a new direct route between peers of different intervals. The line style changes to solid if a route is no longer the newest in subsequent points in time. When establishing a new route, peers add pointers to peers of the other interval, respectively, to their own RT for level $d$. Joining peers will be uniformly distributed among all intervals, which in turn are nodes in a binary tree. Since random trees have logarithmic depth [Kn98], the P2PN scales gracefully and any peer of any interval can be reached in at most $O(\log N)$ hops. An interesting case occurs at time $t_4$ of Fig. 1. Interval $l1$ has already split into $ll2$ and $lr2$ at time $t_3$. Now interval $r1$ splits into $rl2$ and $rr2$ and a direct route is established between both. Since there cannot be more than 4 intervals on level 2, routes between $ll2$ and $rl2$ and $rl2$ and $rr2$ are created. So we make sure the two left intervals and the two right intervals on the same level, respectively, are connected with one another. This ensures load distribution for operations targeted at larger intervals such as $l1$ forwarded by some peer of $rl2$, for instance.

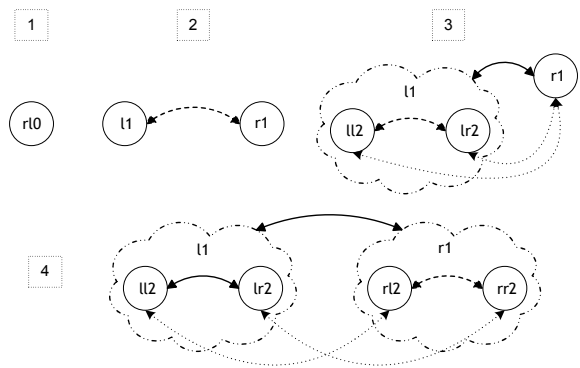

Figure 1: A potential P2PN at times $t_1 \ldots t_4$. Arrows depict links between intervals.

The distributed operation *lookup* finds any datum given its search key. Lookup can start at any one peer and is routed onwards until a peer responsible for the search key is found. We use containedness of keys in intervals to decide on where to route the query next. Consider case 1, the search key falls into the current peer's interval. Then *lookup* returns the datum if it is stored on the peer. If the search key is smaller (greater) than the lower (upper) bound of the current peer's interval, *lookup* will find the smallest interval that only just contains the key. The forwarding peer, then, gets the active virtual connection for this interval by looking it up in its RT and sends the query on to the next peer.

## 4 Coping with fluctuation

As fluctuation may not only degrade performance but, in the worst case, cause P2PNs to fall apart, a general approach to making a P2PN fluctuation insensitive would be desirable. Here we concentrate on mitigation of the impact of fluctuation. We found four consequences of it:

1. data on leaving peers become inaccessible,

2. connections between peers break down if one or more of them leave,

3. a data operation currently processed by the P2PN may die off at a leaving peer,

4. there is a limiting fluctuation rate above which overall performance degrades severely

In contrast to Gummadi et al., who claim that a tree routing geometry offers *no* flexibility in routing a query towards its destination [GGG+03], our tree-based DHT offers solutions to the first three issues. Since every P2PN has its own limiting fluctuation rate, which, once exceeded, renders the system unusable, we think every P2PN must make provisions to deal with the last issue. It remains a future task to quantify the exact rate at which performance of a specific P2PN greatly deteriorates, although Rhea et al. already had a first stab at it [RGRK03].

### 4.1 Data loss

First, to cope with data loss due to leaving peers, we introduce *redundancy through data replication*. If one peer leaves the P2PN this does not pose a threat to the availability of data in general. Peers in our P2PN form data sharing groups of neighboring peers responsible for one interval. If any group member leaves or fails, the remaining peers can still answer queries for data whose search keys belong to their interval. However, the downside of data replication is if the number of peers of an "l" interval, for example, falls below lower limit, $g_c$, the other side, "r", and "l" need to coalesce to form a super interval. Future queries aimed at this interval require that peers of side "l" and "r" replicate their data to each other. This procedure is costly in terms of time, and insert or update operations may need to be deferred until replication ends to avoid inconsistencies. Overall but outlined in a forthcoming paper, our DHT adopts the principle of weak data consistency, i.e., when the P2PN becomes quiescent, data on every peer eventually reach a consistent state.

### 4.2 Failing connections

Second, to cope with failing connections, we introduce *redundancy through backup links*. Every peer needs to maintain at least one connection to another peer per super interval. If this connection fails, the affected peer has to re-establish the connection to a different peer of the same interval. This is time-consuming and may hinder queries from being forwarded to the next peer. Since connection failures are i.i.d. with probability $q_f$, the probability of a failure of all $b \geq 1$ connections (event X), $\Pr\{X\} = q_f^b$ diminishes quickly. Thus, if $l < b$ connections fail, lingering queries can still be forwarded. The P2P system *Kademlia* [MM02], for example, employs so-called $k$-buckets ($k$ is fixed) to achieve the same result. However, the greater the backup factor, $b$, the better for network cohesion but also the higher the maintenance traffic for these additional links. We advocate a connection failure-adaptive mechanism to allow for varying fluctuation rates. So $b$ adjusts to the current situation. Furthermore, peers make sure that links between two intervals are not chosen bidirectionally. This has the positive effect that if a connection fails for some reason, only one peer has to re-establish its connection to the other interval.

### 4.3 Dying off data operations

Third, to cope with peers that are leaving, while one or more data operations are still being processed by them, we introduce the concept of *delivery acknowledgement*. Basically, this concept is about any peer, $\zeta$, sending back a delivery advice to a peer, $\psi$, when it has processed a query and forwarded it on to the next peer. Of course, $\zeta$ only forwards queries if it cannot answer them. $\psi$ waits $t_{\Delta ACK}$ time units for an advice from $\zeta$. If the waiting time is up and $\psi$ did not receive an advice, it reissues the query to a different peer via one of the backup links. In case one such $\zeta$ has in fact failed, this concept makes sure queries eventually reach destination peers. In the other case where $\zeta$ has not failed but may be slow to respond or the network may be congested, duplication of queries occurs. This will eventually lead to duplicate answers reaching the query issuing peer. If some of them are not equal, the peer has to reissue the query to see if the next answer(s) match(es) the larger fraction of the first. Thus the weak consistency principle is met. Choosing timeout $t_{\Delta ACK}$ large enough makes sure duplicate queries crop up rarely.

## 5 Conclusion

Fluctuation in P2PNs poses a major threat to overall network performance. To mitigate this effect, we showed how a tree-based DHT can benefit from rather simple solutions to three general problems pertaining to all P2PNs. Currently, we are building a P2P simulator to quantify the limiting fluctuation rate of our DHT, devise a self-adapting algorithm to keep backup links to a minimum and to explore the impact of varying timeout value $t_{DACK}$.

## References

[FT04] Fahrenholtz, D. and Turau, V.: A tree-based DHT approach to scalable weakly consistent data management. submitted to: *Proc. of the 1st Intl. Workshop on P2P Data Management, Security and Trust*. Zaragoza, Spain. 2004. IEEE Comp. Soc. Press.

[GGG[+]03] Gummadi, K., Gummadi, R., Gribble, S. et al.: The impact of DHT routing geometry on resilience and proximity. In: *Proceedings of the 2003 Special Interest Group on Data Communication*. pp. 381–394. Karlsruhe, Germany. 2003. ACM Press.

[Kn98] Knuth, D. E.: *The Art of Computer Programming. Volume 3: Sorting and Searching*. Addison-Wesley Pub Co. 2nd edition. 1998.

[LSG[+]04] Li, J., Stribling, J., Gil, T. M. et al.: Comparing the performance of distributed hash tables under churn. In: *Proceedings of the 3rd International Workshop on Peer-to-Peer Systems*. San Diego, CA, USA. 2004. Springer-Verlag Heidelberg.

[MM02] Maymounkov, P. and Mazières, D.: Kademlia: A peer-to-peer information system based on the XOR metric. In: *Proceedings of the 1st International Workshop on Peer-to-Peer Systems*. pp. 53 – 65. Cambridge, MA, USA. 2002. Springer-Verlag Heidelberg.

[PF95] Paxson, V. and Floyd, S.: Wide area traffic: The failure of poisson modeling. *IEEE/ACM Transactions on Networking*. 3(3): pp. 226 – 244. 1995.

[RGRK03] Rhea, S., Geels, D. et al.: Handling churn in a DHT. Technical Report UCB/CSD-03-1299. University of California, Berkeley, CA, USA. December 2003.

[SGG02] Saroiu, S. et al.: A measurement study of peer-to-peer file sharing systems. Technical Report UW-CSE-01-06-02. University of Washington, Seatle, WA, USA. 2002.